\documentclass[a4paper]{article}

\usepackage[margin=1in]{geometry} 
\usepackage{amsmath}
\usepackage{float}
\usepackage{amsthm}
\usepackage{amssymb}
\usepackage{graphicx}
\usepackage{xcolor}
\usepackage{tabularx}
\usepackage[ruled,vlined]{algorithm2e}
\usepackage{multirow}
\usepackage{booktabs}
\usepackage[percent]{overpic}  
\usepackage[
    pages=all,
    scale=8,
    color=black,
    opacity=0.05,
    angle=45,
    position={current page.center},
    contents={PREPRINT}
]{background}
\usepackage[utf8]{inputenc}
\usepackage{hyperref}
\hypersetup{unicode,
            hypertexnames=false
            }
\usepackage[numbers,square,compress]{natbib}

\theoremstyle{plain}

\theoremstyle{definition}

\graphicspath{{fig/}}
\usepackage{mathrsfs} 
\usepackage{lipsum}

\title{
Cognitive biases shape the evolution of zero-sum norms
}
\author{Isaak Mengesha$^\dagger$, Meiqi Sun$^\dagger$, Debraj Roy$^\dagger$}

\date{
	$^\dagger$University of Amsterdam \\ 
	\today
}

\begin{document}
	\maketitle
	
\begin{abstract}
Why do maladaptive perceptions and norms—such as zero-sum interpretations of interaction - persist even when they undermine cooperation and investment? We develop a framework where bounded rationality and heterogeneous cognitive biases shape the evolutionary dynamics of norm coordination. Extending evolutionary game theory with bounded rationality and prospect-theoretic utility, we show that subjective evaluation of payoffs systematically alters population-level equilibrium selection, generating stable but inefficient attractors. Counterintuitively, our analysis demonstrates that the benefit of rationality and the cost of risk aversion on welfare behave in non-monotone ways: intermediate rationality enhances coordination, while excessive rationality coupled with strong risk aversion leads to persistent lock-in at low-payoff and zero-sum equilibria. These dynamics produce an endogenous equity–efficiency trade-off: parameter configurations that raise aggregate welfare also increase inequality, while more equal distributions are associated with lower efficiency. In uncertain environments, risk aversion drives societies toward zero-sum worldviews unless loss aversion is strong enough to re-weight cooperative outcomes above the perceived downside risk. These results imply that interventions targeting beliefs directly are unlikely to succeed unless underlying risk environments change, whereas policies that reduce uncertainty or downside risk can shift population-level norm attractors endogenously toward cooperative, positive-sum equilibria.\\

\end{abstract}


\section{Introduction}

A puzzling observation in development economics is the persistence of zero-sum beliefs—the perception that one person's gain necessarily comes at another's expense—in economically disadvantaged environments. These beliefs are empirically widespread, strongly correlated with lower development levels, and demonstrably detrimental to cooperation and investment \citep{rozycka2015belief, carvalho2023zero,liu2024role}. The traditional economic approach would suggest that such maladaptive beliefs should rapidly disappear in competitive environments, as they reduce payoffs and hinder welfare-improving cooperation. While existing psychological explanations offer valuable insights at the individual level \citep{davidai2023psychology, andrews2024zero}, they fail to incorporate evolutionary perspectives that operate at the population and norm level \cite{gavrilets2025evolution}. This omission is particularly significant because payoffs in human interactions are not exogenous—they are subjectively defined and norm-dependent. Utilities are inherently unobservable, but often proxied by monetary outcomes. Even then, monetary payoffs are filtered through beliefs, fairness concerns, and framing effects. People do not simply ``receive" material outcomes; they ``create" subjective value from them \citep{fehr1999theory}. Furthermore, in repeated or networked interactions, actors partly construct or negotiate payoffs through their strategic choices. Their beliefs fundamentally shape these negotiations and therefore influence even the material outcomes \citep{hart2008contracts, geanakoplos1989psychological}. It is thus plausible to assume that beliefs about norms (e.g., `everyone for themselves' or `zero-sumness') will both inform behavior or actions and be reinforced by observed outcomes in a recursive manner \citep{gavrilets2025evolution}.\\

Despite this recursive relationship between beliefs and outcomes, mainstream development economics has largely proceeded in isolation from behavioral insights. The existing literature on poverty and economic growth emphasizes mechanisms such as credit frictions, human capital traps, and structural barriers \citep{azariadis2005poverty, sachs2005end, banerjee2011poor}. Leading accounts in development and growth theory also stress technological change, globalization, and institutional shifts \citep{piketty2014capital, acemoglu2002technology}. Yet much of this work presumes rational responses to constraints and underplays adaptive, path-dependent dynamics \citep{dosi2020rational}. Behavioral studies emphasize individual-level decision-making \cite{gigerenzer2002bounded, barberis2013thirty}, while institutional and network analyses treat payoffs as largely exogenous \cite{nunn2020historical, jackson2007meeting}. What remains missing is a framework that simultaneously integrates bounded rationality, heterogeneous preferences, co-evolving norms, and network dynamics to explain how collective beliefs about the nature of economic interaction—zero-sum versus cooperative—emerge endogenously and persist despite their welfare-reducing consequences. \citep{hoff2016striving}.\\

Evolutionary game theory (EGT) provides the natural foundation for studying endogenous norm formation through selection dynamics.  Classic work explains the emergence of cooperation and persistent payoff asymmetries by modeling how population-level frequencies of strategies evolve when individuals interact repeatedly \citep{axelrod1981evolution, smith1973logic}. However, integration of cognitive biases and beliefs remain limited \citep{nowak2000fairness}. Further, these foundational models usually operate within a critical constraint: the payoff matrix itself is fixed. Agents can vary their strategies (cooperate vs. defect, in-group vs. out-group), but they cannot alter the payoff structure defining which strategies yield which returns. This limitation proves problematic for understanding norm persistence, because norms are precisely about which game society is playing—whether interactions are zero-sum or mutually beneficial, competitive or cooperative, winner-take-all or share-and-share-alike.\\

\begin{figure}[H]
    \centering
    \includegraphics[width=0.8\linewidth]{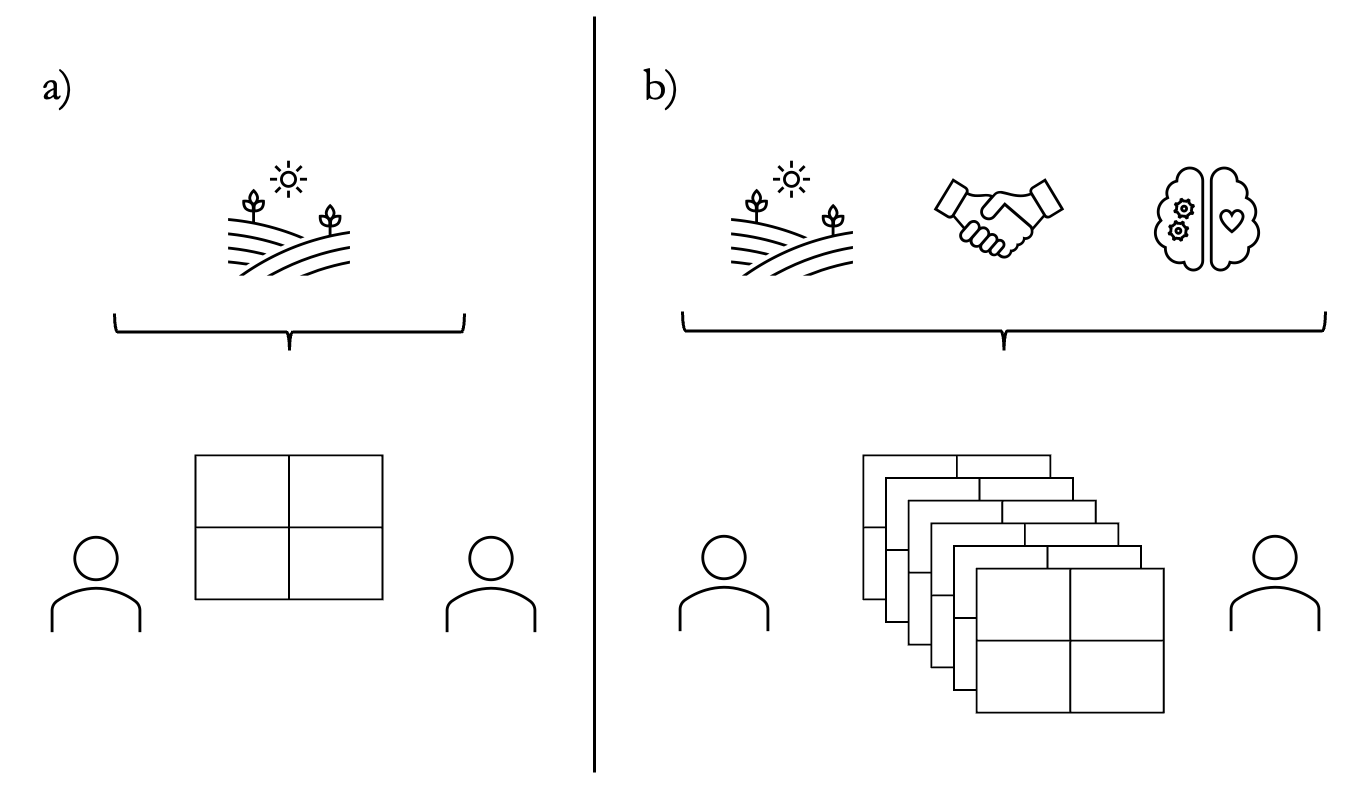}
    \caption{\textbf{Exogenous versus endogenous game determination.} (a) Standard evolutionary game theory assumes a unique mapping from material environment to payoff structure. (b) Our framework accounts for social norms and cognitive heterogeneity, yielding a space of subjectively evaluated games from which population-level selection determines the effective interaction structure.}
    \label{fig:schema}
\end{figure}

To illustrate how the "game" is endogenous rather than fixed, consider three settings. First, farmers sharing an irrigation system can frame their interaction as a Commons Dilemma requiring costly monitoring, or—through trust-building norms—as a Coordination Game where mutual maintenance becomes the default expectation. The physical water remains identical, but the effective payoff structure shifts based on social trust \cite{ostrom1992covenants}. Second, in border disputes, actors may treat territory as indivisible (a Zero-Sum Game where one side's gain is necessarily the other's loss) or engage in issue-linkage to create trade-offs (a Positive-Sum Game). Persistent conflict often reflects a failure to shift the game itself, not merely a failure to cooperate within it \cite{fearon1995rationalist}. Third, employees may view promotions as purely competitive or as shared growth through team success. High risk-aversion can entrench cutthroat "Rat Race" norms even when team-based incentives would yield higher aggregate utility \cite{holmstrom1982moral}. In each case, risk perceptions, loss aversion, and normative expectations actively shape which game the population is playing. As illustrated in Figure \ref{fig:schema}, the set of available games for a given situation is typically larger than one. Our analysis implements the simplest assumption for how agents choose between modes of interaction: selection based on personal welfare. In practice, however, prosocial preferences and other motivational factors will also influence this choice \citep{bierhoff2005prosocial}.\\

If populations were fully rational, maladaptive beliefs should quickly disappear since they reduce payoffs. The persistence of zero-sum norms indicates a systematic deviation in how payoffs are perceived and acted upon, or some form of emergent lock-in at the population level. On the individual level, several mechanisms can anchor zero-sum beliefs by altering the subjective game agents perceive themselves to be playing. Bounded rationality introduces payoff-sensitive noise \citep{mckelvey1995quantal}; prospect-theoretic preferences distort losses and gains; and risk aversion shifts attention toward downside contingencies \citep{tversky1979prospect, barberis2013thirty}. Each of these mechanisms changes agents’ subjective utilities and therefore the norms they find individually optimal. In this sense, individuals do not merely misinterpret a fixed payoff structure—they endogenously reshape the effective game, making zero-sum interaction appear locally rational under uncertainty. Recent evidence suggests that inequality amplifies zero-sum beliefs \citep{davidai2025economic}. This association, however, can arise through two qualitatively different routes. A top-down pathway: observing inequality directly strengthens zero-sum interpretations. Or a bottom-up pathway: an underlying interaction norm—such as competitive behavior—simultaneously generates both inequality and zero-sum beliefs. Distinguishing between these mechanisms is empirically nontrivial, since both produce the same observable correlation. However, why would populations remain locked into low-payoff beliefs -- even across generations. If bounded rationality were the only constraint, learning should reduce zero-sum beliefs; if risk and loss aversion are the cause, increased wealth and security should reduce zero-sum beliefs. Neither mechanism alone suffices to explain norm persistence. Norm-lock in requires -- similar to nash equilibria -- that no individual deviation from the interaction mode can change the norm on the population level.\\

Yet the existing literature treats the relevant components in isolation.
Research on bounded rationality, evolving games, and network structure has developed largely independently. Economic models typically treat interaction rules as fixed, evolutionary models often neglect cognitive distortions, and network models rarely incorporate heterogeneity. This compartmentalization obscures the interaction effects that are central to explaining norm formation in complex social systems. Prior work has explored how preferences evolve through cultural selection \citep{Alger2013, WangWu2023} and how institutions reshape the feasible set of interactions \citep{Bowles1998, frey2020dynamic}. There has been no work to our knowledge on how cognitive heterogeneity—specifically bounded rationality and prospect-theoretic evaluation—generates endogenous game selection even while holding material environments fixed.
To address this gap, we treat the concept of norms as ``population-level attractors in the space of possible games". In Fig. \ref{fig:zero} we see how this space is disaggregated into well-known families of games, that the population navigates. Fixed points occur when populations become locked into particular strategic interactions (games) that resist incremental change through individual adaptation. In such situations, small groups or individuals cannot unilaterally shift the prevailing norm without incurring substantial costs, even when the norm generates suboptimal aggregate outcomes. Under this framework, zero-sum beliefs emerge not because individuals are irrational, but because zero-sum games become evolutionarily stable. The novelty of our approach lies in jointly considering the following mechanisms: behavioral mechanisms (e.g. bounded rationality), evolutionary game theory (e.g. replicator dynamics) and network dynamics (e.g. homophilic rewiring) -- to investigate interaction effects between them.\\

Our analysis makes three principal contributions. First, we demonstrate that equilibrium selection depends critically on agents' subjective evaluation of payoffs, showing that cognitive processes fundamentally shape which norms emerge. Populations with identical material payoffs but different utility functions (risk-neutral vs. loss-averse) converge to entirely different equilibrium—one to cooperative coordination games, another to competitive Prisoner's Dilemma. This demonstrates that cognitive heterogeneity is not a second-order perturbation but a primary determinant of which norms emerge. Second, we establish that norm trajectories are shaped by robustness across diverse environments, explaining why societies with similar initial conditions can diverge dramatically. Counterintuitively, we find that intermediate rationality ($\lambda \approx 2.5$) maximizes cooperation and welfare, while excessive precision paired with strong risk aversion generates persistent lock-in at low-payoff and zero-sum equilibria. This non-monotone effect arises because excessive rationality causes brittle optimization within each game rather than robust coordination across the population's ecological mixture of games. Third, we show that the equity-efficiency trade-off emerges endogenously from these adaptive processes, with configurations that maximize welfare systematically increasing inequality. This trade-off arises because high rationality allows cognitively precise agents to extract larger payoffs from less precise partners. Conversely, configurations that equalize payoffs do so by constraining everyone to near-random play, sacrificing efficiency. To quantify these effects, we work with a ``zero-sumness index" (see Fig. \ref{fig:zero}) based on the correlation of payoffs between players, characterizing how different equilibria foster cooperative versus competitive interpretations of interaction.

\section{Methods}
To operationalise this framework, we integrate three elements: (1) Quantal response equilibrium to capture bounded rationality, (2) prospect-theoretic utility to embed heterogeneous risk preferences, and (3) evolutionary game dynamics on endogenous game space to study norm formation. This integration allows us to trace how populations with different distributions of cognitive parameters (precision $f_\lambda$, loss aversion $f_\eta$) converge to qualitatively different norm attractors. By doing so, we provide a behavioral-evolutionary micro-foundation for understanding how individual-level cognitive biases systematically shape population-level norm emergence, and ultimately, why societies with similar initial conditions can diverge into persistent traps.

\subsection{Game space}
To study norm evolution across the full space of strategic interactions, we require a unified parametrization that preserves strategic content under affine payoff transformations. Following \cite{hauert2002effects}, we adopt the $U{\text-}V$-parametrization, which maps all symmetric $2{\times}2$ games with canonical payoffs $(R,S,T,P)$ to a two-dimensional plane while maintaining equivalence classes under rescaling. Following the $U{\text-}V$ parametrization, we map the row payoff matrix to
\begin{equation}
    \text{Game}= g =
    \begin{bmatrix}
    1 & U\\[2pt]
    V & 0
    \end{bmatrix},\qquad
    U=\frac{S-P}{R-P},\ \ V=\frac{T-P}{R-P},
\end{equation}
so that affine transformations of payoffs leave strategic content invariant. We exclude the degenerate case $R{=}P$. Unless otherwise noted we consider $(U,V)\in[-1,2]^2$, which covers the standard families of $2{\times}2$ games within a common scale; the mapping situates classic regimes such as Prisoner’s Dilemma, Stag Hunt, and Hawk–Dove in distinct regions of the $(U,V)$ plane \cite{mengesha2025evolutionary}. Furthermore, each game in the plane is normalized by subtracting the mean $\mu_g$ of the possible payoffs from all entries.

\begin{figure}[H]
\centering
\begin{tabularx}{\textwidth}{@{}>{\centering\arraybackslash}m{0.44\textwidth}@{\hspace{1em}}>{\raggedright\arraybackslash}m{0.52\textwidth}@{}}

\includegraphics[width=\linewidth]{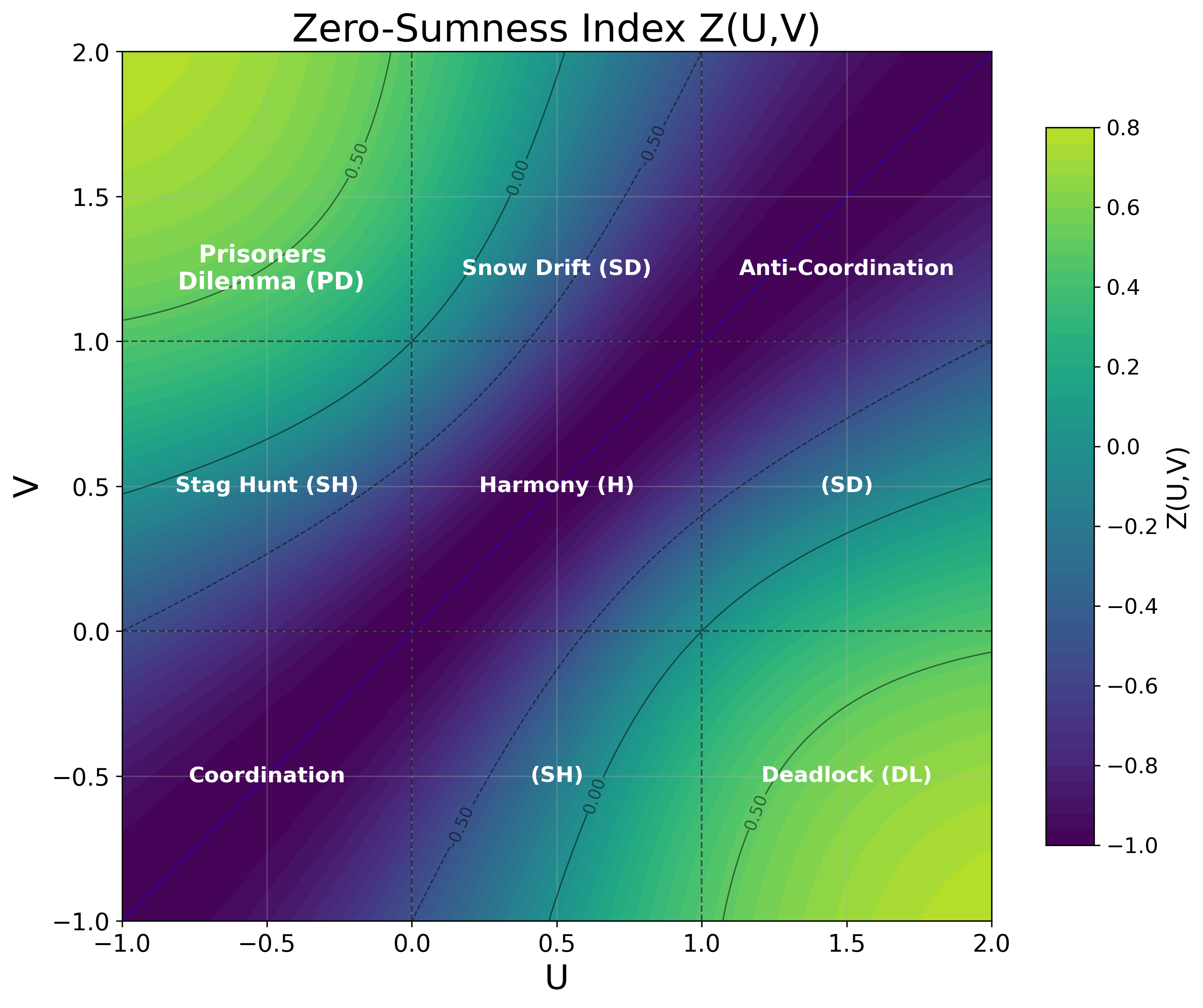} &

\begin{minipage}[c]{\linewidth}
\begin{equation}
    \pi =
\begin{array}{c|cc}
     & C & D \\
\hline
C & (1,1) & (U,V) \\
D & (V,U) & (0,0)
\end{array}
\end{equation}
\vspace{1cm}
\begin{equation}
Z(U,V) \;=\; -\mathrm{corr}\!\left(\pi^{\text{row}},\,\pi^{\text{col}}\right),
\end{equation}
\end{minipage}
\end{tabularx}

\caption{Qualifying the degree of zero-sumnes for the UV-plane. None of the games can be fully zero sum ($Z=1$), however they can approach full mutual interests ($Z=-1$). Game types form expectations about how "aligned" mutual interactions are. Consequently it shapes the beliefs about cooperation and trust.}

\label{fig:zero}
\end{figure}

\subsection{Evaluation of games}
 Agents evaluate material payoffs through subjective utility functions that capture cognitive biases. We investigate two specifications to isolate distinct mechanisms. Prospect-theoretic utility (${U}_1$) incorporates reference dependence and loss aversion. Exponential utility (${U}_2$) captures risk aversion without reference dependence. We formalize the two possible design choices as shown below:

\begin{equation}
\begin{aligned}
\mathcal{U}_1(c \mid r,\eta) &=
\begin{cases}
(c-r)^{\alpha(\eta)}, & c \ge r,\\[4pt]
-\omega\,(r-c)^{\beta(\eta)}, & c < r,
\end{cases} \\[8pt]
\mathcal{U}_2(c \mid \eta) &= -\exp\!\bigl(-\eta c\bigr) + \eta c + 1,
\end{aligned}
\end{equation}\\

where $\mathcal{U}_1$ denotes prospect-theoretic utility with reference point $r$, curvature $\alpha(\eta),\beta(\eta)$ depending on control parameter $\eta$ and loss-aversion parameter $\omega$ , and $\mathcal{U}_2$ denotes the linex form, with $\eta$ governing the direction and magnitude of asymmetry. This specification isolates the effect of concave utility, separately from loss aversion. Strategy choice follows a logit quantal response equilibrium (LQRE). For agent $i$, strategy probabilities take the Boltzmann form
\begin{equation}
p_i(s)\;\propto\;\exp\{\lambda_i\,\mathcal U_i(s)\}, \qquad 
Z_i=\sum_{k}\exp\{\lambda_i\,\mathcal U_i(s_k)\}, \quad
p_i(s)=\frac{\exp\{\lambda_i\,\mathcal U_i(s)\}}{Z_i},
\end{equation}
with partition function $Z_i$ ensuring normalization. Options with equal $\mathcal U_i$ receive equal mass; probabilities depend only on utility differences. $\lambda_i$ can be interpreted as rationality and tunes precision from random choice $(\lambda_i\to 0)$ to near best response $(\lambda_i\gg 1)$. In the limit of large $\lambda$ we expect the strategies to converge to at least one of the possible many nash equilibria (NE).
We assume heterogeneity exists in $(\lambda,\eta)$. We denote by $\pi(g;\lambda,\eta)$ the expected payoff to type $(\lambda,\eta)$ in game $g$. Let $f_\lambda(\lambda)$ and $f_\eta(\eta)$ denote the distributions of heterogeneity parameters:

\begin{equation}
f_\lambda(\lambda) \sim \text{LogNormal}(1,\sigma^2), 
\quad f_\eta(\eta) \sim  \text{LogNormal}(1.4, \sigma_\eta^2).
\end{equation}

Furthermore, let $p_i(g;\lambda,\eta)\in(0,1)$ denote the logit–QRE probability of playing strategy~1 in game $g$ after mapping material payoffs through $\mathcal U$ as above using LQRE.  For a given $(\lambda,\eta)$, this requires solving the coupled fixed points of two LQRE best responses. No closed-form solution exists.
\subsubsection*{Population-Level Aggregation}
Population-level behavior needs to account for the heterogeneity in $(\lambda,\eta)$, which govern the sharpness of the responses to payoff differences and the way risk is evaluated. The relevant observable is the average probability in game $g$ of picking strategy one,
\[
S(g)=\mathbb{E}[p_1(g;f_\lambda,f_\eta)]
=\int_0^\infty\!\!\int_0^\infty p_1(g;\lambda,\eta)\,f_\lambda(\lambda)f_\eta(\eta)\,d\lambda\,d\eta,
\]
which aggregates heterogeneous boundedly rational responses into a single $S(g)\in[0,1]$. For computation, both $\lambda$ and $\eta$ are lognormal distributed and $p_1(g;\lambda,\eta)$ is smooth and bounded. This permits efficient evaluation of $S(g)$ via Gauss–Hermite quadrature after transforming to standard normal variables, yielding accurate approximation with only a small number of QRE evaluations per game. This expectation transforms the complex landscape of individual bounded-rational responses into a single aggregate strategy $S(g) \in$  characterizing population behavior. Games with higher $S(g)$ generate more cooperative play on average, while those with lower values tend toward defection.

\subsubsection*{Fitness Landscape Construction}
We define the fitness of a game type $g$ as its expected material payoff against the population. When a type-$g$ player meets a type-$g'$ player, their expected payoff is
\[
K(g,g') = S(g)S(g')\cdot 1 + S(g)(1-S(g'))\cdot U + (1-S(g))S(g')\cdot V.
\]
Uniform random matching over the space $\mathcal G=[U_{\min},U_{\max}]\times[V_{\min},V_{\max}]$ implies
\[
\Phi(g) = \tfrac{1}{|\mathcal G|}\int_{\mathcal G} K(g,g')\,dg',
\]
with discrete approximation $\Phi(g) = \tfrac{1}{N}\sum_{i,j} K(g,(U_i,V_j))$ on a grid. This construction parallels the standard payoff-to-fitness mapping in evolutionary models but extends it to heterogeneous, boundedly rational agents through the population response function $S(g)$.

\subsubsection*{Attractor Identification}
Game dynamics are governed by the fitness gradient $\dot g=\nabla_g\Phi(g)$, analogous to adaptive dynamics in evolutionary biology and replicator dynamics in economics. Attractors correspond to locally stable game types $g^*$, characterized by vanishing gradient $\nabla_{U,V}\Phi(g^*)=0$ and negative definite Hessian $\nabla^2_{U,V}\Phi(g^*)\prec 0$. These points mark evolutionary steady states of the strategic landscape under our bounded rationality framework. We interpret the evolution of game parameters $(U, V)$ as shifts in how populations collectively interpret and enforce their interactions, not as changes to underlying material reality -- as all payoff matrices are normalized.

\subsection{Validation via Agent-Based Model}
While mean-field approximations (Eq. 5) characterize the asymptotic stability of norms in infinite populations, they fundamentally miss the dynamics path dependence central to our argument. In finite populations with heterogeneous agents, the transition between norms is driven by stochasticity —rare events where a cluster of agents successfully coordinates on a risky but superior norm. An analytical equilibrium analysis would identify stable points but cannot predict which basin of attraction a population will likely fall into given its history and network structure. \\

To assess the robustness of the analytical framework, we implement an agent-based model (ABM) that mirrors the structure of the mean-field setting but allows for localized interaction, endogenous game choice, and network coevolution \cite{mengesha2025evolutionary}. The ABM serves as a micro-foundation check: it tests whether the population-level attractors derived analytically also emerge when individual agents with heterogeneous cognitive limits interact directly. Simulations use $N{=}200$ agents over $T{=}200$ periods, replicated ten times for averaging. Agents are initialized with heterogeneous $(\eta_i,\lambda_i)$ and game parameters $(U_i,V_i)$, with wealth and reference points starting at zero. A five-period warm-up without belief adjustment, learning, or rewiring prevents early-history artifacts. Each subsequent period agents draw opponents from their network, update strategies via logit QRE given their precision $\lambda_i$, play and accumulate payoffs, choose new games using a softmax rule over utilities, and adjust network links by income homophily. This design captures the joint evolution of strategies, preferences, and interaction structure in a decentralized environment.\\

The ABM allows validation along two key dimensions. First, we track the distribution of wealth $W_{i,t}$ to study inequality dynamics. Second, we monitor the convergence of strategies and game choices $(U,V)$, comparing the emergent fixed points to the analytically predicted attractors. Results confirm that the mean-field attractors remain stable under agent-level dynamics, while the ABM highlights additional dispersion and path-dependence arising from heterogeneity and network evolution. Full implementation details and pseudo-code are reported in the Appendix section \ref{aproxalgo}.


\section{Results}
\subsection{Aversions shape the emergence of norms}

We first establish that equilibrium selection depends fundamentally on how agents evaluate payoffs. Figure 2 compares fitness landscapes under three utility specifications: risk-neutral (panel a), risk-averse without loss aversion (panel b), and prospect-theoretic with loss aversion (panel c).  We interpret the dominant game present in the population as a social norm. Note that due to normalization, there is no \textit{a priori} reason to prefer one game over another. Consequently, the norms are endogenous behavioral fixed points on a population level determined by interaction and preferences of individuals. First, we investigate the impact of changes to the utility function to the normative fixed points, by introducing risk aversion and loss aversion respectively. To compute these fixed points, we compare the fitness landscapes generated using our mean-field approach and validate this approximation against agent-based simulations. Second, we examine the robustness of these attractors to heterogeneity in rationality and risk attitudes. This two-step analysis highlights both the functional-form dependence of attractors and the secondary role of characteristics on a population-level.\\

\begin{figure}[H]
    \centering
    \includegraphics[width=0.9\textwidth]{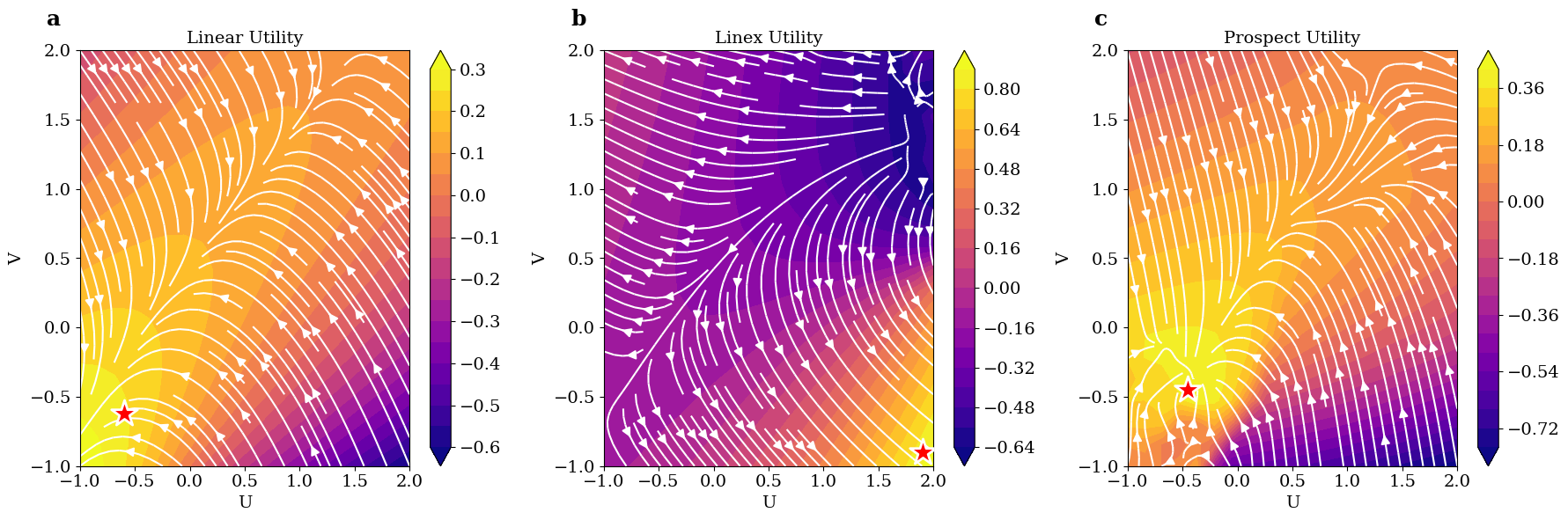}
    \caption{We investigate the sensitivity of fixed points to the utility function. Using our mean-field approach to estimate fitness of games/behavior. \textbf{(a)} We find that rational (as in not risk averse/seeking) population gravitates towards pure coordination games (see fig. a). \textbf{(b)} Once we introduce risk aversion by applying $\mathcal{U}_2$, we notice clear fixed points emerge. Those are particularly focused on the diagonal corners (PD and DL). \textbf{(c)} Finally we investigate the fixed points of prospect utility $\mathcal{U}_1$ and find it to be located in the coordination game regime. The red star indicates the result of the ABM simulation with dynamic network properties. }
    \label{fig:fixedpoint}
\end{figure}

We distinguish between three scenarios: no risk or loss-aversion, risk-aversion only, and both risk and loss-aversion. As we investigate figure \ref{fig:fixedpoint}, we notice distinct shapes between the resulting fitness landscapes. The 'rational' population has an attractor in pure coordination games. This is a consequence of normalization concentrating payoff in the first row/column and maximizing payoffs (Fig. \ref{fig:fixedpoint}a). Introducing aversions in (Fig. \ref{fig:fixedpoint}b) and (Fig. \ref{fig:fixedpoint}c), we notice the emergence of different attractors. By introducing risk aversion only (b), the population converges towards either PD or DL. Both are games with defection (D) as NE strategy, where PD yields a Pareto inefficient strategy and DL a Pareto efficient one. This means for populations evolving on this landscape, there is a risk of them getting temporarily `trapped' in the Pareto inefficient PL norm. The location of these attractors is a consequence of the strictly increasing transformation of payoffs by the Linex utility function. It leaves the ordinal rankings unchanged and placing weight on risk-dominating strategies. Norms that are robust to any strategy deployed by the counter party are preferred. The category of coordination games exhibits a high sensitivity and is therefore unattractive, ruling out the main diagonal. In $UV$-space the payoff gradient always points diagonally ($U\uparrow,V\downarrow$ or $U\downarrow,V\uparrow$), ensuring convergence to PD/Deadlock. \\

Adding loss-aversion changes the normative fixed point towards coordination games (Fig. \ref{fig:fixedpoint}c), where both the pareto optimal strategy and NE is cooperation (C). Loss aversion introduces a reference point, meaning payoffs are not evaluated in absolute terms or using a globally concave transform. Instead losses -- compared to that reference point -- loom larger than equally large gains. When facing uncertainty, strategies are shaped increasingly by the goal of avoiding losses. Consequently, this replaces risk dominating strategies with loss minimizing ones, which are found in games of coordination and harmony. Interestingly, adding risk-aversion increases defection and the presence of zero-sum norms, adding loss-aversion on top increases the presence of cooperation and the absence of zero-sum norms. 


\subsection{Norms and cognitive biases coevolve}
We learned that the choice of utility function determines the qualitative nature of norm-attractors on the population level. However, the question arises: are the observed attractors robust to changes in the distribution of agent traits —such as those documented across societies \cite{falk2018global} — or do they systematically shift when varying populations? We find that the distribution of agents' rationality and risk aversion meaningfully influences the norms present. As a key mechanism, we need to distinguish between \textit{within-game} and \textit{cross-game} optimality. Within a given game or norm, agents earn payoffs by playing best responses to the anticipated strategies of others, which can be approximated by QRE. In contrast, when evaluating games in a heterogeneous ecology, opponents may face different payoff structures, so their incentives need not be similar. This creates adversarial pressure that favors security strategies such as maximin or minimax, which safeguard against worst-case losses rather than "best-response" to anticipated behavior. Only in the special case of zero-sum games do Nash equilibria and security strategies coincide. Outside of this limit $Z<1$, the gap between best response and security reasoning makes outcomes sensitive to heterogeneity in risk preferences and bounded rationality. \\

\begin{figure}[H]
    \centering
    \includegraphics[width=0.85\textwidth]{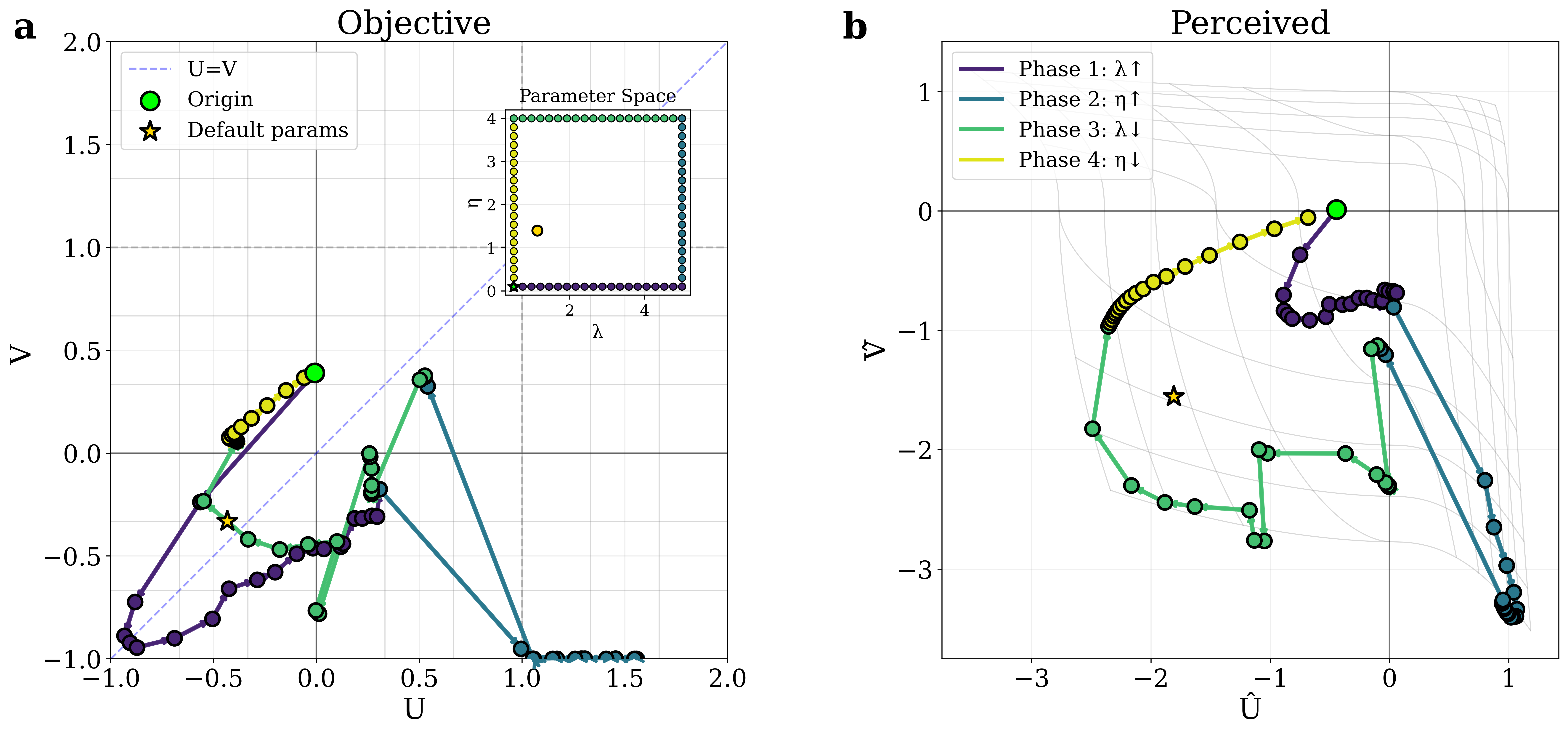}
    \caption{ We investigate how varying the mean of $\lambda$ or $\eta$ on the population level affects the fixed points in game space. \textbf{(a)} Following a closed loop in parameter space, we observe the fixed point move mostly along the coordination diagonal. However, for high rationality and risk aversion, the system makes a jump to DL. \textbf{(b)} visualizes the distortions stemming from the utility function for the same trajectory. In this subjective game space, agents never enter zero-sum games but stay within coordination or assurance games.  }
    \label{fig:sensitivity_fixedp}
\end{figure}

The experiment of Fig. \ref{fig:sensitivity_fixedp}a follows a closed loop in the parameter space of risk aversion and rationality $(\eta,\lambda)$ and identify the corresponding fixed points. Here, we vary the means $(\Bar{\mu},\Bar{\lambda})$ of the respective LogNormal distributions. The starting point of the curve lies in $(0.1,0.5)$  where agents’ behavior is largely characterized by noise, placing the fixed point near the average over the UV-plane. As rationality increases while risk aversion remains low, the fixed point shifts toward coordination games. Coordination first maximizes within-game payoffs by concentrating positive value in $C$ for both players. This constitutes a local optimum. As rationality increases further, the population becomes increasingly able to identify the ecology global optimum and approaches the more (\textit{cross-game}) robust harmony game via SH.\\ 

As we increase risk aversion $\eta$, the preference shifts towards risk dominant strategies. As a consequence the attractor shifts towards the new local optimum, Deadlock games. Note that at high values for both $\eta$ and $\lambda$, prospect-theoretic and linex attractors converge. However, with increasing risk aversion losses become so distorted, that risk dominant strategies \textit{within-game} (DL) become maladaptive. Hence, the attractor shifts back to the globally robust norm of harmony games. Following a similar reasoning, we observe a half-circle (green) via SH towards coordination games and returning back towards SH in the direction of harmony games. We reach the origin, as we reduce risk aversion and completely return to harmony games. However, in reality, agents do not optimize for payoff but for their subjective utility. The distortions introduced result in an ”effective” U-V plane that the system traverses (Fig. \ref{fig:sensitivity_fixedp}b). Note that the majority of the trajectory, occurs in quadrant $(\hat{U},\hat{V})$ that corresponds to coordination norms. Interestingly, agents perceive none of the norm attractors from the class of defection dominant DL/PD norms. Fig. \ref{fig:sensitivity_fixedp}b depicts a representative distortion for $\eta=1.4$. There can be no fixed mapping over the entire trajectory, as the transformation is dependent on the varied parameter $\eta$.\\

Taken together, the trajectory analysis shows that the fitness of norms cannot be reduced to within-game payoffs; it depends on stability across heterogeneous interactions. Rather than assuming population-wide agreement, we observe that agents adapt locally and stable attractors emerge. Analogous to Nash equilibrium within a game, norms can be understood as fixed points of interaction rules where unilateral deviation offers no improvement. This framing highlights how cognitive biases in risk perception and bounded rationality coevolve with social norms, shaping which conventions remain evolutionarily stable.


\subsection{Endogenous Equity-Efficiency Trade-off}
We have investigated the emergence and location of population-level fixed points. However, if said fixed-point have no meaningful influence on wealth or welfare, this insight remains fairly inconsequential. In the following, we will investigate how parameter variation affects population-level wealth. First, we investigate how aggregate performance depends on the distribution of agent-level cognitive traits. For this we utilitise the ABM that does not rely on the assumption of a well-mixed population. Instead we introduce an underlying network with homophilic rewiring to capture the fact of clustering behaviour in social network across traits. For a more detailed description of the model consult section \ref{ABM} in the appendix.\\

\begin{figure}[H]
    \centering
    \begin{minipage}{0.48\textwidth}
        \centering
        \begin{overpic}[width=\linewidth]{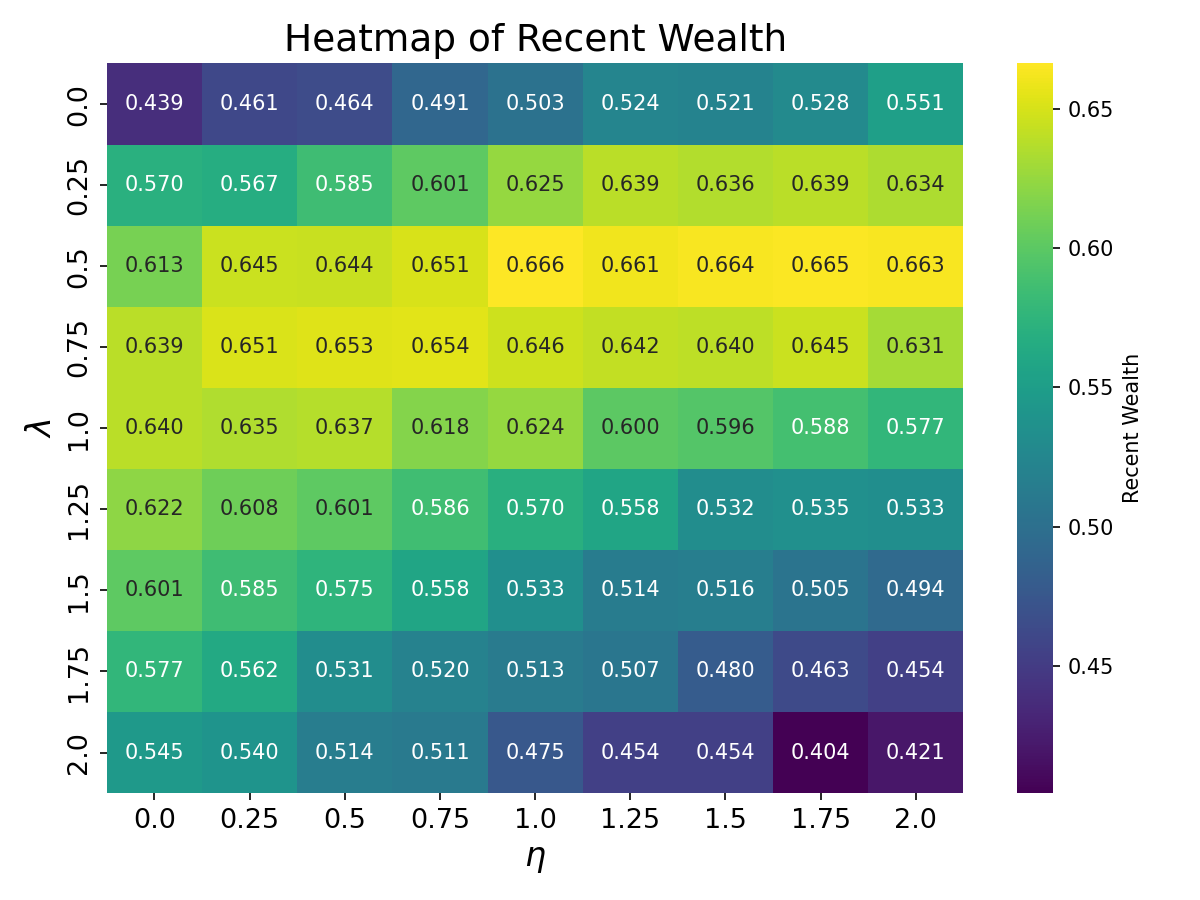}
            \put(1,75){\textbf{a}} 
        \end{overpic}
    \end{minipage}
    \hfill
    \begin{minipage}{0.48\textwidth}
        \centering
        \begin{overpic}[width=\linewidth]{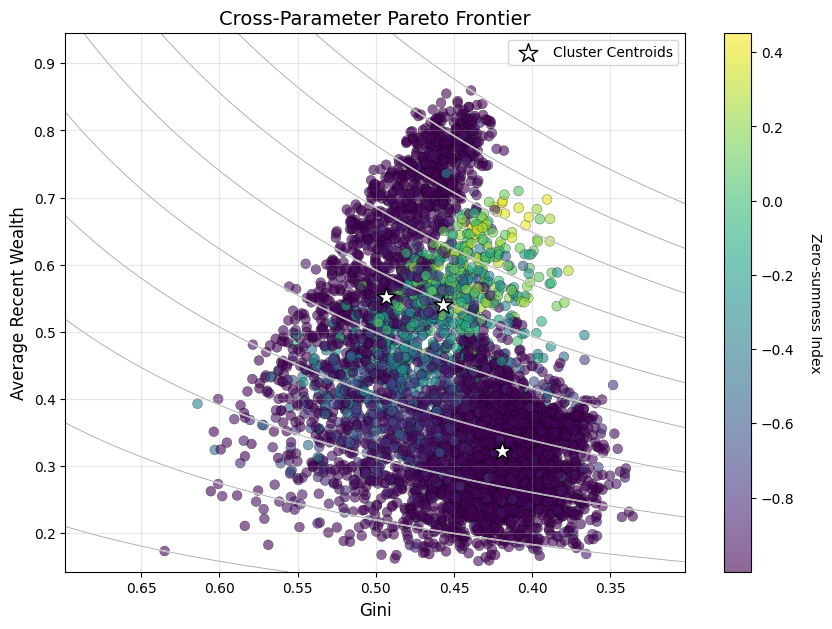}
            \put(1,75){\textbf{b}}
        \end{overpic}
    \end{minipage}
    \caption{Sensitivity of welfare to population level characteristics and system parameters. \textbf{(a)} We shift the distribution of population values $\eta$ and $\lambda$ by a fixed constant and measure the average population income. Notice the inversion of the influence of risk aversion with increasing rationality. \textbf{(b)} Here we vary system parameters in line with the sensitivity analysis and visualize each configuration in the tuple of average income and gini coefficient. Furthermore we color each point according the fixed point games zero summness index. Welfare isoclines are computed using Sen welfare index \cite{sen1976real}.}
    \label{fig:heterogeneity}
\end{figure}

Neoclassical benchmarks would suggest that higher precision $\lambda$ monotonically increases payoffs by moving play closer to the Nash equilibrium, while higher risk aversion $\eta$ lowers expected payoffs by reducing exposure but enhances stability. Our results complicate this view. Population-level income and cooperation do not rise monotonically with $\lambda$ but instead follow an inverted-U shape (see Fig. \ref{fig:heterogeneity}a). Beyond an intermediate level of precision, further approximation of Nash equilibrium ceases to increase payoffs. This plateau emerges because in heterogeneous populations, excessive precision locks agents into exploitative strategies that in some cases reduce robustness against a plurality of norms, echoing critiques that hyper-rational adjustment can undermine system performance in disequilibrium contexts \cite{dosi2020rational}. Furthermore, risk aversion produces additional non-linear effects. At low precision, increasing $\eta$ is associated with higher rather than lower payoffs, as loss-averse behavior promotes early coordination across most games. Once the population crosses a threshold in $\lambda$, the relationship inverts, aligning with conventional intuition that excessive caution reduces returns. Note that inversion of the benefits of risk aversion on a population level is also related to the corresponding norm-attractor in game-space. Meaning the respective performance of a given parameter combination ($(\eta, \lambda)$ actively shapes the norms in which the agents operate.

Next, we explore how outcomes vary across parameter configurations. Since the model lacks a priori “correct” values for $(\lambda,\eta)$, we conduct sensitivity analysis (Section \ref{sensitivity} in Appendix) and examine how welfare and inequality respond. We compute welfare following Sen’s (1976) formulation as $W = \overline{W^R}(1 - \alpha G)$, where $\overline{W^R}$ denotes average recent wealth, $G$ is the Gini coefficient, and $\alpha=1$ the inequality aversion parameter \cite{sen1976real}. This reveals a systematic tension: configurations that maximize total payoffs also generate higher inequality, whereas more equal wealth distributions coincide with lower aggregate welfare. The resulting Pareto frontier (Fig.~\ref{fig:heterogeneity}b) suggests that the system cannot simultaneously realize highest welfare and lowest inequality. Unlike standard treatments where the equity-efficiency trade-off reflects technological constraints or policy choices, here the frontier emerges from the interaction of cognition, preferences, and network dynamics. Configurations that raise aggregate welfare do so by enabling high-precision agents to exploit strategic opportunities, necessarily increasing inequality. Conversely, configurations that equalize payoffs do so by locking in suboptimal norms. We identify three clusters (white star as centroids). The properties of these clusters can be found in the section \ref{clusters} in the Appendix.  They mainly differ in population level rationality: Increased rationality boosts welfare but make outcomes sensitive to opponent mix and network position, increasing inequality; reducing rationality equalizes payoffs at the cost of efficiency. Interestingly, we find a third cluster at a similar welfare level to the high rationality cluster, where zero-sum norms are dominant. This implies a triangle, where high welfare either demands high cognitive capacity, or lower trust (ie. zero-sum norms) or welfare is traded of for more equality.  \\

The system as a whole displays the characteristics of a complex adaptive system, in which outcomes arise endogenously from the interaction of boundedly rational agents embedded in a co-evolving network. This immediately complicates attribution: higher-order interactions make it difficult to cleanly separate the role of structural versus cognitive parameters, echoing long-standing challenges in the inequality literature where emergent properties resist decomposition into first-order effects \cite{jackson2007meeting,dosi2020rational}. To make progress, we first establish convergence for the norm /game level, for given parameter distributions of agents. Next we proceeded with testing the stability of the underlying network structure, which ensures that subsequent distributional results are not artifacts of transient topology. We then isolate contributions of the two main components — network topology and agent heterogeneity — on wealth distribution (See Appendix section \ref{distributive}). The network primarily shapes the variance and persistence of inequality, generating heterogeneous outcomes across agents without predetermining who benefits. In this sense, it conditions the opportunity structure but does not fix the allocation of advantage, consistent with network-based accounts of inequality emphasizing access and diffusion rather than initial endowments \cite{kossinets2006empirical}. By contrast, heterogeneity in cognitive precision and risk preferences directly reduces aggregate welfare. Even when the mean level of cognition is identical in both simulations variance in $\lambda$ and $\eta$ produces friction and strategic mismatch, lowering total payoffs. At the same time, it determines who occupies the upper tail of the distribution: agents with higher precision are disproportionately represented in high-wealth positions (see Appendix, Fig.~\ref{fig:rat-wealth-corr}), consistent with theories of cumulative advantage in adaptive environments \cite{weibull1997evolutionary}. These observations hold robustly for both income and accumulated wealth.  \\

\section{Discussion}
Carvalho \textit{et al.} show that zero-sum beliefs evolve in zero-sum environments \cite{carvalho2023zero}. We demonstrate that the perception of the environment matters for the development of such beliefs and norms as well. While it is well established that prospect theory affects individual choices in games, we show that it also shapes population-level equilibrium selection. Combined with QRE, our approach captures the core dimensions of bounded rationality and provides a methodological advance by endogenizing norm formation.


\subsection{From Individual Biases to Population Lock-In}
Our results highlight three mechanisms through which bounded rationality and preference heterogeneity shape the evolution of norms and welfare outcomes. Each connects to and extends existing literature, but in ways that shift the explanatory emphasis.\\

First, utility functions shape norm equilibrium selection. Classical accounts of equilibrium selection stress noise and stochastic perturbations \cite{carlsson1993global,kandori1993learning}. Our findings demonstrate that the functional form of utility itself is decisive: prospect-theoretic reference points can flip risk dominance, while linex transformations preserve it. This implies that norm-equilibrium selection is not simply a property of the payoff matrix but of the mapping from payoffs to subjective utilities. The result resonates with cross-cultural experiments showing systematic variation in cooperation across societies and correlations with belief in zero-sum interactions \cite{henrich2001cooperation,falk2018global}. Norm diversity, therefore, can be traced not only to differences in material incentives or historical shocks but also to how payoffs are internally evaluated.\\

Second, precision and risk aversion are not monotone in welfare. The inverted-U relationship we find between rationality and payoffs challenges the neoclassical presumption that more rationality strictly improves outcomes. At intermediate levels, higher precision aids coordination, but beyond that threshold, agents become locked into brittle, exploitative strategies. This echoes critiques of hyper-rational models in complex adaptive systems, where attempts at perfect optimization destabilize adjustment \cite{dosi2020rational,guzman2020towards}. Logit equilibrium analyses similarly show that the welfare implications of noise are non-linear \cite{goeree2001logit}. Strategies that are locally optimal in specific games---such as pure coordination games---under perform when agents face an ecology of heterogeneous games. This aligns with meta-game approaches that emphasize competition among institutional forms \cite{frey2020dynamic,alger2013homo}. From an evolutionary standpoint, the analogy is to ecological stability: robustness requires adaptability across multiple environments rather than fine-tuned performance in a single one \cite{hofbauer1998evolutionary}. This frames the emergence of social norms in line with nash equilibria -- behavioral fixed points that no subgroup can unilateraly change, without incurring cost. This helps explain why societies can lock into divergent institutional trajectories: their institutions differ in cross-context robustness, not only in immediate efficiency.\\

Finally, the equity-efficiency frontier is endogenous. The trade-off between aggregate welfare and inequality emerges endogenously from the coevolution of cognition, norms, and networks. Surprisingly, we find that that low-trust (zero-sum) norms can yield high welfare outcomes. Suggesting there exist conditions zero-sum beliefs might be adaptive. However, our model neglects potential negative externalities of low trust environments by focusing on exclusively pair-wise interactions. Overall these considerations contrast with standard treatments that view the equity-efficiency frontier as a technological or policy constraint. Instead, the frontier itself shifts with the underlying adaptive processes. Network structure generates persistent heterogeneity in exposure and opportunities, while preference diversity amplifies inequality through selection effects \cite{jackson2019inequality,weibull1997evolutionary}. Interventions therefore do not merely redistribute along a fixed frontier but can reshape the frontier itself by altering the dynamics of norm formation and diffusion.\\

Taken together, these findings suggest that welfare and inequality should be understood not merely as outcomes of endowments or policies but as endogenous byproduct of cognitive limits, evolutionary mechanisms, and network dynamics. We show that zero-sum norms can emerge as a fixed point in populations of high risk aversion, which might be appropriate for environments of high volatility. This hints that global variations in zero-sum beliefs may reflect accurate equilibrium conditions, rather than maladaptive distortions. This shifts attention from comparative statistics to evolutionary pathways and from exogenous constraints to endogenous trade-offs, as an alternative mechanism for the unaccounted for variance in risk preferences across individuals and societies \cite{guiso2008risk}.\\

\subsection{Policy implication}
Our findings carry non-obvious policy implications. Standard development interventions aim to change beliefs directly—through education campaigns, social marketing, or incentivized cooperation experiments \cite{WDR2015}. Our model predicts such interventions will fail if they don't address the underlying risk environment in line with tentative evidence \cite{hoff2016striving}. Consider two stylized interventions:\\

\textbf{Belief-targeting}: Persuade individuals that interactions are positive-sum, shifting subjective game perceptions. In our model, this corresponds to exogenously moving agents' $(U_i,V_i)$ toward coordination games. However, as long as the population-level attractor remains in the PD/DL region (due to unchanged distribution in $\eta$ or $\omega$), individual deviations will be selected against. Agents who cooperate based on shifted beliefs will be exploited by the majority playing risk-dominant strategies, reinforcing the zero-sum equilibrium. \textbf{Risk-targeting}: Reduce environmental volatility through insurance, safety nets, or income stabilization programs. This lowers optimal $\eta$, shifting the population-level fitness landscape (Fig. 2). Once $\eta$ falls below the threshold, the attractor itself moves from PD/DL to Stag Hunt or Harmony regions, endogenously changing beliefs as a byproduct. Positive-sum norms become individually rational, and zero-sum beliefs dissipate through evolutionary selection. This suggests targeting risk environments is more effective than targeting beliefs, a prediction testable in established field experiments on insurance provision combined with measures of zero-sum beliefs over time \cite{delavallade2015managing}.\\

More generally, our model generates three testable hypotheses: First, higher risk aversion should be associated with stronger zero-sum beliefs, as greater sensitivity to uncertainty makes individuals more likely to perceive social interaction as conflictual rather than mutually beneficial. Second, greater loss aversion should weaken zero-sum beliefs, since an asymmetric response to losses fosters coordination when cooperation reduces exposure to risk. Third, cooperation and welfare should be maximized at intermediate levels of rationality, while excessive precision leads to coordination failure and fragile equilibria. There are findings on the sensitivity of inequality to zero-sum norms in line with our findings, but can be considered weak evidence at best \cite{davidai2025economic}. We leave empirical validation to future work, as available cross-country data suffer from aggregation issues and small-N limitations unsuitable for the within-country micro-mechanisms our model describes. Our model can now serve as sandbox that can guide empiricial investigation on the coevolutions of aversions and belief in zero-sum games.

\section{Acknowledgments}

The author gratefully acknowledges institutional support from the University of Amsterdam. DR acknowledge the support from the Netherlands eScience project of the Dutch NWO, under contract 27020G08, titled “Computing societal dynamics of climate change adaptation in cities”.

\bibliographystyle{unsrtnat} 
\bibliography{refs}

\newpage
\renewcommand{\thesubsection}{\Roman{subsection}}
\setcounter{subsection}{0}
\setcounter{page}{1}
\section*{Appendix}
\setcounter{section}{0}
\subsection{Sensitivity Analysis} \label{sensitivity}

Global sensitivity analysis examined five parameters ($\alpha$, $\lambda$, 
$N$, $\eta$, $\omega$) using PAWN and Sobol methods. We used Saltelli sampling with N=512 base samples, generating 3584 parameter 
combinations. Each combination is replicated three times, with each simulation running 200 agents over 200 time steps. PAWN 
indices measured distributional sensitivity, while Sobol indices 
quantified variance contributions. Statistical significance was assessed via bootstrap at 10\%, 5\%, and 1\% levels.

\begin{table}[H]
\centering
\begin{tabular}{lccccc}
\toprule
y & $\alpha$ & $\lambda$ & $N$ & $\eta$ & $\omega$ \\
\midrule
Gini & 0.12* & 0.03* & 0.42* & 0.04 & 0.10* \\
Recent Wealth & 0.04 & 0.03* & 0.52 & 0.04* & 0.12 \\
Zerosum & 0.07 & 0.04* & 0.21* & 0.06* & 0.09* \\
\bottomrule
\end{tabular}
\caption{Mean sensitivity indices for different dependent variables (PAWN method). Significance levels: $\dagger$ 10\%, $\ddagger$ 5\%, * 1\%}
\label{tab:pawn_sensitivity}
\end{table}

Table~\ref{tab:pawn_sensitivity} presents the PAWN sensitivity indices, revealing distinct patterns of parameter influence across different outcomes. For Gini coefficient, the normalization parameter $N$ demonstrates the strongest sensitivity ($0.42, p<0.01$), indicating the primary driver effect of inequality dynamics in the model. The homophily parameter $\alpha$ ($0.12, p<0.01$), loss aversion parameter $\omega$ ($0.10, p<0.01$) and rationality parameter $\lambda$ ($0.03, p<0.01$) show significant but varying degree effects. Interestingly, the risk aversion parameter $\eta$ shows no significant effect on inequality. The Recent Wealth metric displays a different sensitivity profile. $N$ again emerges as the dominant factor, exerting even stronger influence than on inequality, while $\alpha$ and $\omega$ demonstrate non-significant influences on agents' income. For the Zerosumness metric, which captures the game attributes possessed by the agent, all parameters except $\alpha$ show statistically significant effects. The normalization parameter $N$ remains important, though its influence is weaker than for the other outputs. $\omega$, $\eta$ and $\lambda$ all contribute significantly to determining the zero-sum attribute of agents. Overall, the PAWN analysis highlights that the normalization parameter consistently exerts the strongest influence across all three output metrics. The differential sensitivity patterns across outputs indicate that parameters governing agent behavior ($\eta$, $\lambda$, $\omega$, $\alpha$) have outcome-specific effects, with their importance varying depending on whether we examine inequality, income or game attributes.

\begin{table}[H]
\centering
\small
\begin{tabular}{lcccccc}
\toprule
\multirow{1}{*}{} & \multicolumn{5}{c}{Sobol Indices ($S_T$($S1$))} \\
\cmidrule(lr){2-6}
 & $\alpha$ & $\lambda$ & $N$ & $\eta$ & $\omega$ \\
\midrule
Gini & 0.29 (0.20) & 0.08 (0.00) & 0.77 (0.67) & 0.10 (0.03) & 0.19 (0.04) \\
Recent Wealth & 0.03 (0.00) & 0.03 (0.00) & 0.98 (0.93) & 0.03 (0.00) & 0.06 (0.01) \\
Zerosumness & 0.53 (0.00) & 0.49 (0.01) & 0.88 (0.41) & 0.55 (0.03) & 0.59 (0.03) \\
\bottomrule
\end{tabular}
\caption{Sobol sensitivity indices for model outputs. Values shown as $S1$ / $S_T$, where $S1$ represents first-order effects and $S_T$ represents total-order effects.}
\label{tab:sobol_sensitivity}
\end{table}

To complement the PAWN analysis, we conducted a Sobol variance-based sensitivity analysis (Table~\ref{tab:sobol_sensitivity}). The results corroborate the PAWN findings, with the normalization parameter $N$ consistently dominating across all outputs. parameter N consistently dominating across all outputs. Notably, the Sobol analysis reveals more substantial parameter effects for Zerosumness compared to PAWN results. All five parameters show moderate-to-strong total-order indices ($S_T>0.48$), with $\alpha$ (0.534), $\eta$ (0.553) and $\omega$ (0.585) and $\lambda$ (0.486) demonstrating cpmparable importance to each other. This suggests significant interaction effects among parameters when determining competitive dynamics. For Gini coefficient and Recent Wealth, both methods align alosely: $N$ dominates, while other parameters show moderate effects. The convergence between PAWN and Sobol analyses for these outputs strengthens confidence in the robustness of our sensitivity findings.

\subsection{Specification of corresponding ABM} \label{ABM}

\paragraph{Agents.}
Each agent $i$ carries fixed cognitive parameters and evolving balance-sheet variables:
risk sensitivity $\eta_i$, bounded-rationality precision $\lambda_i$, cumulative wealth $W_{i,t}$, and recent wealth $W^R_{i,t}$. Wealth is the running sum of realized material payoffs, with a floor at zero if the cumulative value would become negative; we do not model interest or discounting. Recent wealth aggregates the last five realized payoffs,
\begin{equation}
    W^R_{i,t}=\frac{1}{n_t}\sum_{j=0}^{n_t-1} P_{i,t-j},\qquad n_t=\min\{t+1,5\},
\end{equation}
so that during the first five periods $W^R_{i,t}$ is the average over available history. Both $W_{i,t}$ and $W^R_{i,t}$ update after each game and feed forward into decision-making and linking. 

Intrinsic heterogeneity is initialized as $\eta_i\sim\mathrm{Log}\mathcal N(1.4,0.5)$, $\omega_i\sim\mathrm{Log}\mathcal N(2,0.5)$  and $\lambda_i\sim\mathrm{Log}\mathcal N(1,0.5)$, yielding a right-skewed distribution of risk sensitivity and decision precision. These parameters remain fixed throughout simulations; only payoffs, strategies, wealth, and network ties adapt. Each agent also holds an idiosyncratic payoff specification $(U_i,V_i)$ defining their current row matrix $M_i=\bigl(\begin{smallmatrix}1&U_i\\[2pt] V_i&0\end{smallmatrix}\bigr)$, which evolves via belief adjustment and learning below. Optional payoff normalization by the sum of matrix entries can be enabled before play; unless stated, we use the normalized $U$–$V$ form. 

Material payoffs are evaluated relative to a dynamic reference point given by $r=W^R_{i,t}$ via a prospect-theoretic utility $\mathcal U(c\mid W^R,\eta_i)$,
\begin{equation}
\begin{aligned}
\mathcal U(c\mid W^R,\eta) &=
\begin{cases}
(c-W^R)^{\alpha(\eta)}, & c \ge W^R,\\[2pt]
-\omega\,(W^R-c)^{\beta(\eta)}, & c < W^R,
\end{cases} \\[6pt]
\alpha(\eta) = \max\!\Bigl\{0.2,\ \tfrac{1}{1+\eta}\Bigr\},& \quad
\beta(\eta)  = \max\!\Bigl\{0.2,\ \tfrac{1}{1+0.5\,\eta}\Bigr\}.
\end{aligned}
\end{equation}

This single-parameterization of curvature and loss aversion bounds elasticities away from zero, ties both domains to $\eta$, and uses recent wealth as the moving reference. 

\paragraph{Playing and Learning}
Agents could have partial information about opponents’ payoffs and temporarily adjust their own $U$ and $V$ toward the opponent’s before a game. Let $\delta_b\!\ge\!0$ denote payoff dependence. If $i$ faces $k$, then, prior to play,
\begin{equation}
U'_i=(1-\delta_b)\,U_i+\delta_b\,\frac{U_i+U_k}{2},\qquad
V'_i=(1-\delta_b)\,V_i+\delta_b\,\frac{V_i+V_k}{2}.
\end{equation}
While we explore this parameter the default setting is set as no shared information ($\delta_b=0$), with agents believing to play identical games. After playing and receiving payoffs, $i$ may update her \emph{persistent} payoff specification toward the opponent’s via a probabilistic learning gate derived from a psychometric function with sensitivity $\lambda_i$ and stimulus $(u_k-u_i)$, the difference in prospect utilities realized in the match:
\begin{equation}
p_i(\text{learn})=\frac{1}{1+\exp\big(-\lambda_i\,(u_k-u_i)\big)}.
\end{equation}
Conditional on learning, $i$ applies a convex update with learning rate $\delta_g\in[0,1]$,
\begin{equation}
U_{i,t+1}=(1-\delta_g)\,U_{i,t}+\delta_g\,U_{k,t},\qquad
V_{i,t+1}=(1-\delta_g)\,V_{i,t}+\delta_g\,V_{k,t}.
\end{equation}
Thus $\delta_g=0$ yields full retention, $\delta_g=1$ copies the opponent’s matrix. Learning is unilateral per encounter; the initiator compares utilities and potentially updates, while the opponent’s persistent $(U,V)$ remains unchanged that round. 

\paragraph{Decision rules.}
Strategy choice follows a logit quantal response equilibrium (LQRE) over prospect-valued expected payoffs. For agent $i$,
\begin{equation}
P_i(s)\;=\;\frac{\exp\{\lambda_i\,\tilde U_i(s)\}}{\sum_{k}\exp\{\lambda_i\,\tilde U_i(s_k)\}},
\end{equation}
where $\tilde U_i(s)$ denotes the prospect-utility expectation of pure strategy $s$ given the current opponent’s mixed strategy and $i$’s match-specific payoff matrix. In two-player, two-strategy games, cooperation probabilities $(P^1_C,P^2_C)$ solve the coupled fixed point
\begin{align}
P^1_C&=\frac{\exp\{\lambda_1\,[P^2_C U^1_{CC}+(1-P^2_C)U^1_{CD}]\}}{\exp\{\lambda_1\,[P^2_C U^1_{CC}+(1-P^2_C)U^1_{CD}]\}+\exp\{\lambda_1\,[P^2_C U^1_{DC}+(1-P^2_C)U^1_{DD}]\}},\\
P^2_C&=\frac{\exp\{\lambda_2\,[P^1_C U^2_{CC}+(1-P^1_C)U^2_{DC}]\}}{\exp\{\lambda_2\,[P^1_C U^2_{CC}+(1-P^1_C)U^2_{DC}]\}+\exp\{\lambda_2\,[P^1_C U^2_{CD}+(1-P^1_C)U^2_{DD}]\}},
\end{align}
with $P^i_D{=}1{-}P^i_C$. We compute the LQRE fixed point numerically every interaction. $\lambda_i$ tunes precision from random choice ($\lambda_i{\to}0$) to near best response ($\lambda_i >> 1$ ). 

\paragraph{Network dynamics.}
We initialize the social graph as one of Erd\H{o}s--R\'enyi, Watts--Strogatz, Holme--Kim configurations. After play, links adapt by income homophily. For any potential link $(i,k)$ we use the logistic attachment function
\begin{equation}
P_{\mathrm{con}}(i,k)=\frac{1}{1+\exp\{\alpha\,(|W^R_i-W^R_k|-\rho)\}},
\end{equation}
with homophily strength $\alpha$ and threshold $\rho$ (default $\rho{=}2$). Existing links are severed with complementary probability $P_{\mathrm{cut}}=1-P_{\mathrm{con}}$, making dissimilar pairs more likely to dissolve. To preserve the expected average degree, deletions without replacement trigger random additions, and additions without deletion trigger random removals elsewhere. New links are considered primarily among second-order neighbors to reflect triadic closure under homophily.

\subsubsection*{Algorithm of ABM}
Unless otherwise stated, simulations use $N{=}200$ agents and $T{=}200$ periods and are replicated ten times for averaging. Initialization draws $\eta_i$, $\lambda_i$, and $(U_i,V_i)$ as above; wealth terms start at $W_{i,0}=W^R_{i,0}=0$. We include a five-period warm-up during which no belief adjustment, learning, or rewiring occurs to avoid early-history dominance. 

\begin{algorithm}[H]
\caption{Detailed simulation steps for each agent. Every model step all agents complete these operations.}
\label{tab:simulation-steps}
\KwIn{Parameters $\{\delta, \lambda, \eta, W\}$ for every agent $i \in N$}

\For{period $t$}{
    \For{each agent $i$}{
        \begin{enumerate}
            \item \textbf{Selection:} sample random opponent from neighborhood;
            \item \textbf{Adjustment:} update payoffs via $\delta$;
            \item \textbf{Strategy:} update action with Logit QRE($\lambda_i$);
            \item \textbf{Play Game:} update $W_i, W_i^R$;
            \item \textbf{Choose Game:} softmax over $U(\eta_i; W_i^R)$;
            \item \textbf{Network Update:} rewire by income-homophily;
        \end{enumerate}
    }
}
\end{algorithm}

Steps 2, 5, and 7 are inactive for $t<5$ (warm-up). Mutation introduces controlled noise while keeping exploration subcritical by scaling $P_m$ with $N^{-2}$. We track inequality via the distribution of $W_{i,t}$ and dynamics via convergence of $(U_i,V_i)$ and strategy profiles. We track (i) inequality via the wealth distribution and (ii) system learning dynamics via adaptation rates of strategies/games.

\subsection{Algorithm for Aproximation} \label{aproxalgo}

\subsubsection*{Computation via Gaussian Quadrature}
Direct evaluation of $S(g)$ requires repeatedly solving the QRE for many draws of $(\lambda,\eta)$. Since both parameters are lognormally distributed and the QRE solution $p_1(g;\lambda,\eta)$ is smooth and bounded, Gauss–Hermite quadrature provides a more efficient approach than Monte Carlo. The lognormal can be reparametrized in terms of standard normal variables:
\[
\lambda=\exp(0.5\xi/\sqrt{2}),\ \xi\sim\mathcal N(0,1),\qquad
\eta=\exp(\sigma_\eta\zeta/\sqrt{2}+\mu_\eta),\ \zeta\sim\mathcal N(0,1).
\]
This yields
\[
S(g)=\tfrac{1}{\pi}\!\int_{-\infty}^{\infty}\!\!\int_{-\infty}^{\infty}
p_1\!\left(g;\exp(0.5\xi/\sqrt{2}),\,\exp(\sigma_\eta\zeta/\sqrt{2}+\mu_\eta)\right)
e^{-\xi^2}e^{-\zeta^2}\,d\xi\,d\zeta,
\]
which can be approximated using $n$-point Gauss–Hermite quadrature:
\[
S(g)\approx\tfrac{1}{\pi}\sum_{i=1}^n\sum_{j=1}^n
w_i w_j\,p_1\!\left(g;\exp(0.5\xi_i/\sqrt{2}),\,\exp(\sigma_\eta\zeta_j/\sqrt{2}+\mu_\eta)\right).
\]
In practice $n{=}5$ achieves the same accuracy as Monte Carlo with about $M{=}1000$ draws, reducing cost from $O(|G|^2 M)$ to $O(|G|^2 n^2)$—a $\sim40\times$ speedup.

\begin{algorithm}[H]
\caption{Approximation of Attractors in $(U,V)$-space}
\KwIn{Numerical grid $\{(U_i,V_j)\}$, heterogeneity distributions $f_\lambda,f_\eta$, threshold $\epsilon$}
\KwOut{Estimated attractors in $(U,V)$-space}

\For{each grid point $(U_i,V_j)$}{
    \tcp{Step 1: Population strategies}
    Compute $S(U_i,V_j)=\mathbb{E}_{\lambda,\eta}[p((U_i,V_j);\lambda,\eta)]$ 
    using Gauss--Hermite quadrature or Monte Carlo sampling\;
}

\For{each grid point $(U_i,V_j)$}{
    \tcp{Step 2: Fitness landscape}
    Evaluate $\Phi(U_i,V_j)=\sum_{(U',V')} K((U_i,V_j),(U',V'))\,\rho(U',V')$\;
}

\tcp{Step 3: Gradient field}
Approximate partial derivatives $(\partial\Phi/\partial U,\;\partial\Phi/\partial V)$ 
at each grid point using finite differences\;

\tcp{Step 4: Attractors}
Identify local maxima as grid points $(U_i,V_j)$ where 
$|\nabla \Phi(U_i,V_j)| < \epsilon$ and $\Phi(U_i,V_j)$ exceeds neighboring values\;

\Return{Set of attractors in $(U,V)$-space}
\end{algorithm}

\subsection*{Formal payoff orderings for Game Classification}

The classification in Fig.~\ref{fig:zero} is drawn in terms of the parameters
$(U,V,0,1)$. These correspond directly to the canonical payoffs of a symmetric
$2\times 2$ game: 
\[
U \mapsto S \; (\text{Sucker}), \quad 
V \mapsto T \; (\text{Temptation}), \quad
0 \mapsto P \; (\text{Punishment}), \quad 
1 \mapsto R \; (\text{Reward}).
\]

Each game type is defined by a characteristic ordering of these four payoffs:

\begin{itemize}
    \item \textbf{Prisoner’s Dilemma (PD):} $T > R > P > S$.
    Defection strictly dominates cooperation, but mutual cooperation is better than mutual defection.

    \item \textbf{Deadlock (DL):} $T > P > R > S$.
    Defection dominates and also yields the socially preferred outcome.

    \item \textbf{Snowdrift (SD):} $T > R > S > P$.
    Symmetric in $U$ and $V$. Each player prefers to free-ride if the other cooperates, but prefers cooperation over mutual defection.

    \item \textbf{Stag Hunt (SH):} $R > T > P > S$.
    Symmetric in $U$ and $V$. Mutual cooperation is payoff-dominant, but defection is risk-dominant when trust is low.

    \item \textbf{Coordination (C):} $R > P > T,S$.
    Multiple equilibria exist, favoring convergence on the same action.

    \item \textbf{Anti-Coordination (AC):} $T,S > R,P$.
    Players are best off choosing opposite actions.

    \item \textbf{Harmony (H):} $R > S > T > P$.
    Cooperation strictly dominates defection.
\end{itemize}

Thus each region in the $(U,V)$ plane corresponds to one of these canonical
payoff orderings. However there still exist some game classes not captured by our formalization, as there exist restrictions on symmetry (diagonal payoffs are always identical between row and column players) and parameter range.

\subsection{Distributive properties and cooperation} \label{distributive}

\begin{figure}[H]
    \centering
    \includegraphics[width=0.9\textwidth]{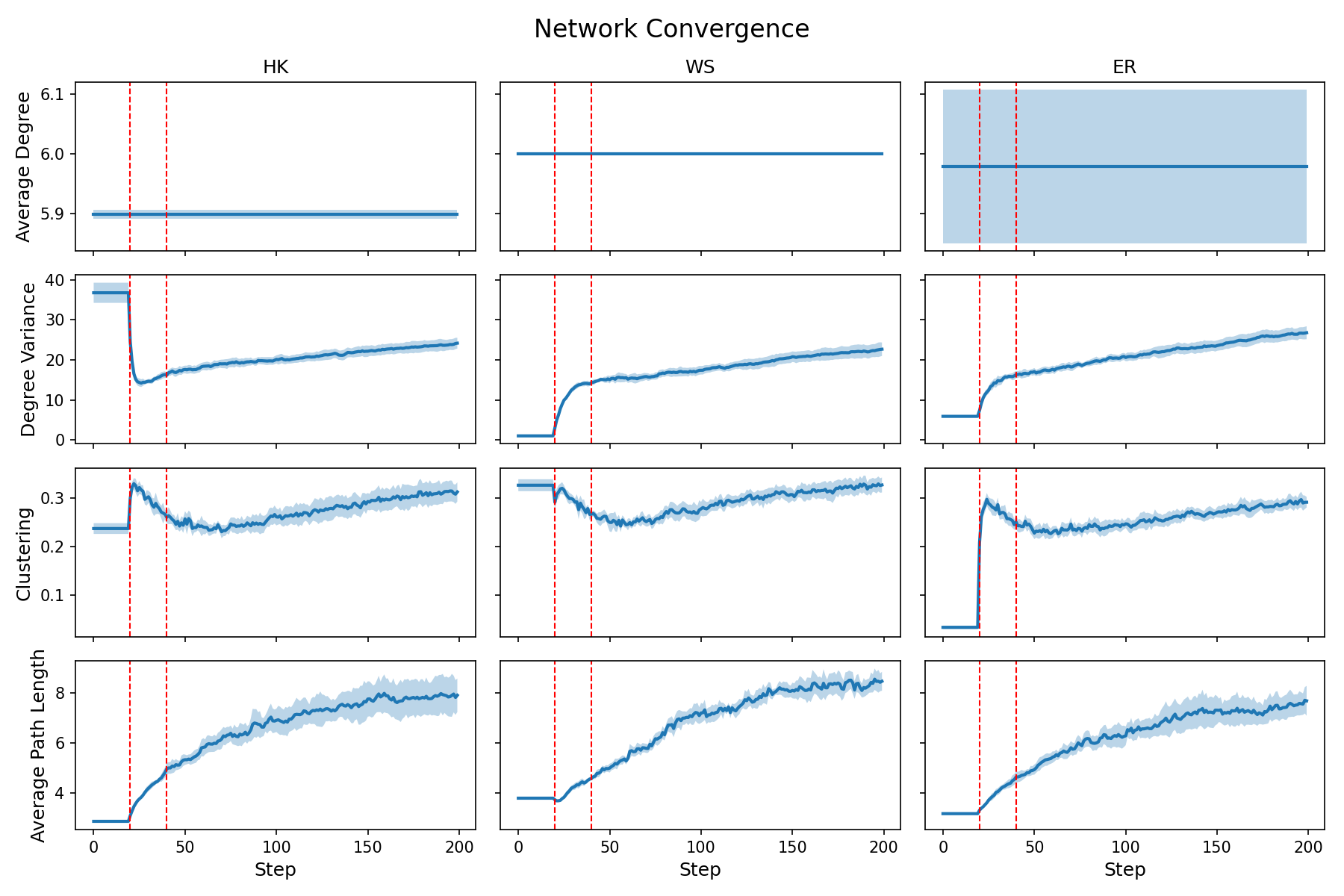}
    \caption{We explore the convergence to the steady state network structure for different initial network topologies. They share a common steady state.  }
    \label{fig:net-convergence}
\end{figure}

\begin{figure}[ht]
    \centering
    \includegraphics[width=0.45\textwidth]{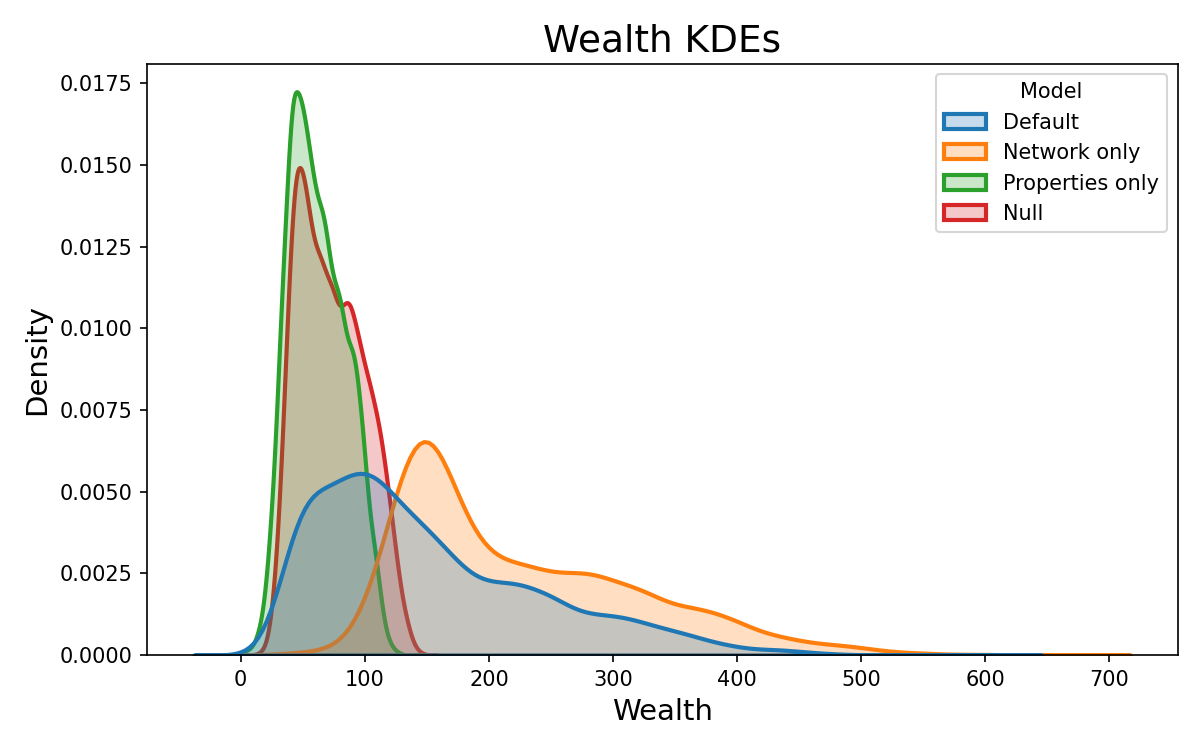}
    \hfill
    \includegraphics[width=0.45\textwidth]{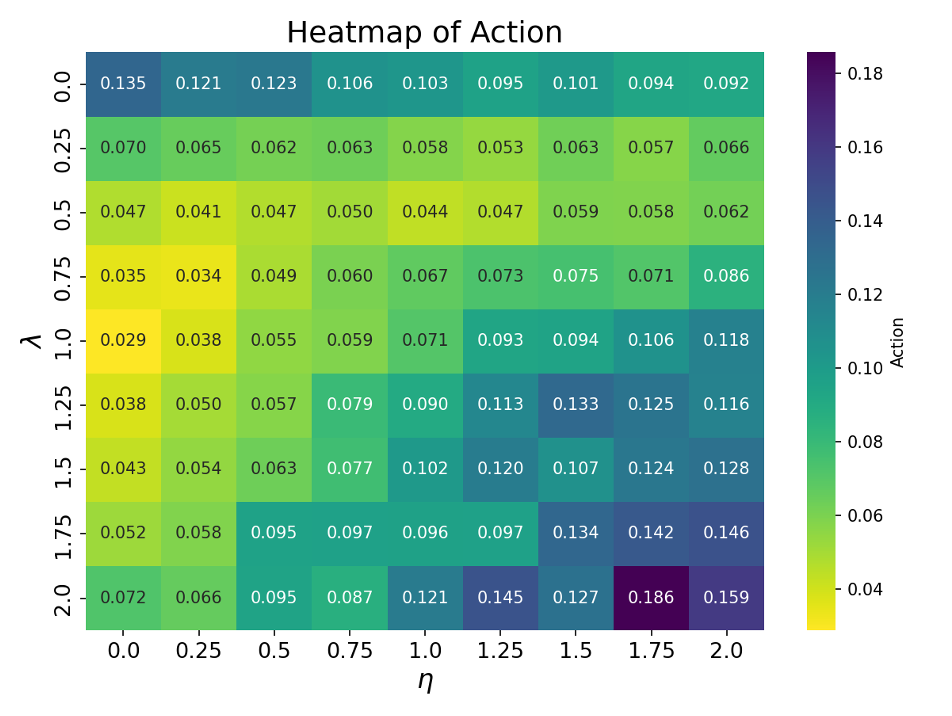}
    \caption{Distribution of recent wealth and total wealth across different model configurations.}
    \label{fig:distri}
\end{figure}

Note, that while the overall shape for Income and Cooperation in figure \ref{fig:distri} are similar, they have non-identical maxima. With cooperation peaking for medium precision and zero risk aversion, income peaks for medium precision and substantial risk aversion.

\subsection{Properties of Clusters}\label{clusters}

\begin{figure}[H]
    \centering
    \includegraphics[width=0.8\textwidth]{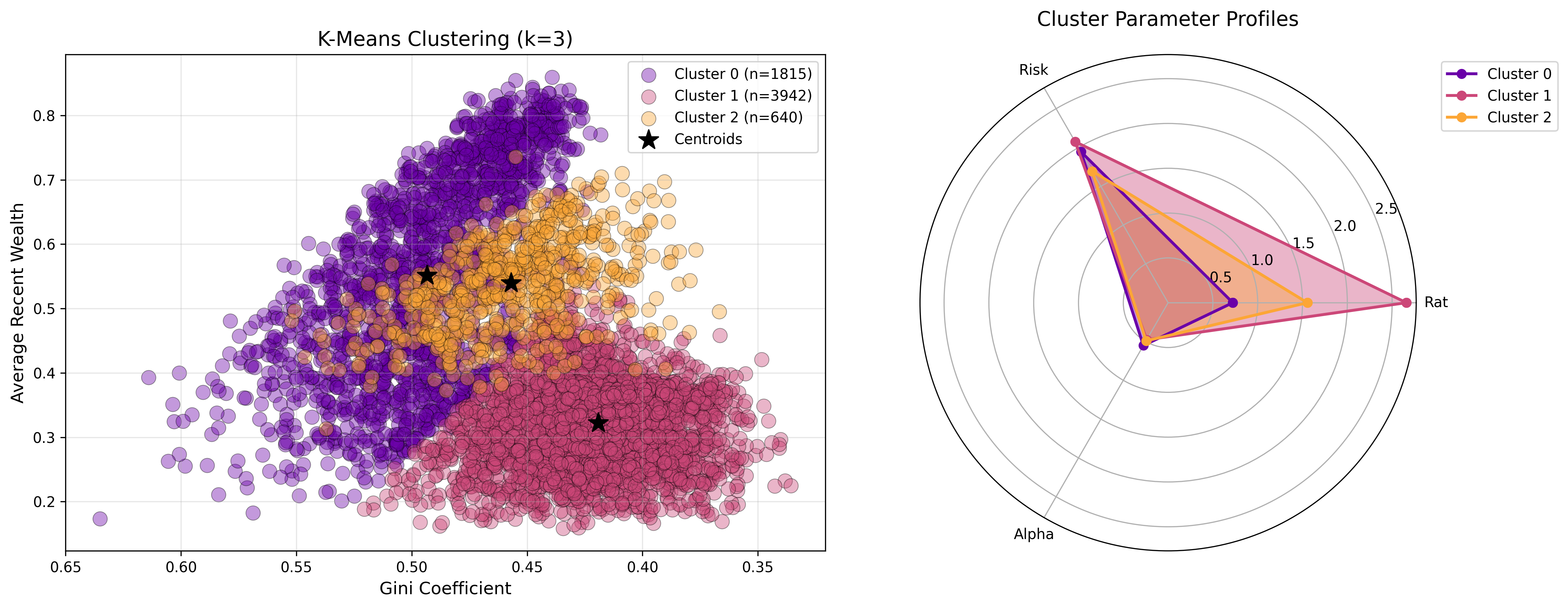}
    \caption{Investigating the properties of clusters formed from GSA simulations. They mostly differ in rationality, where increases in rationality increase equality but decrease income.}
    \label{fig:clusteringprops}
\end{figure}


\subsection{Correlations}
\begin{figure}[H]
    \centering
    \includegraphics[width=0.48\textwidth]{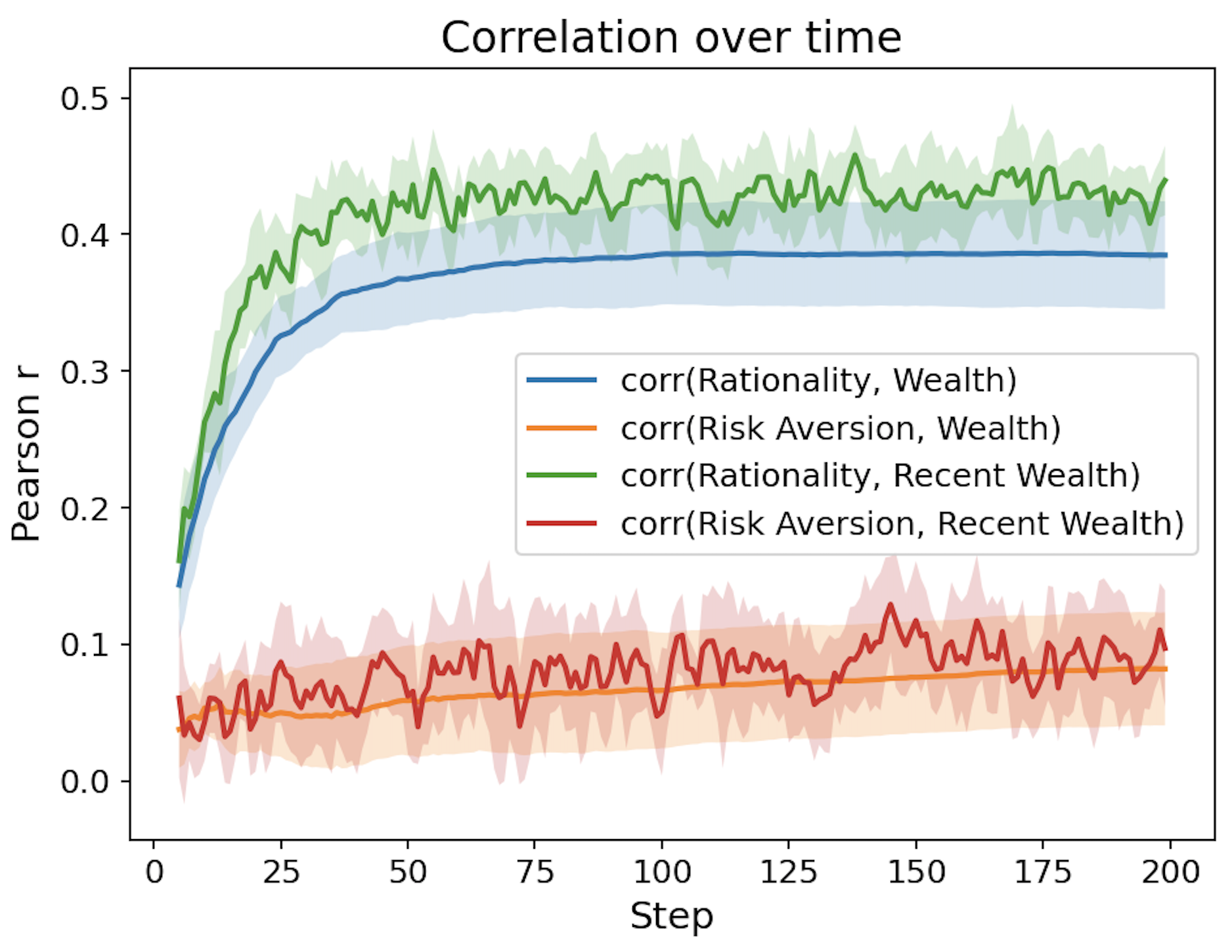}
    \caption{Correlation of agents risk aversion and rationality with wealth and income.}
    \label{fig:rat-wealth-corr}
\end{figure}

Figure~\ref{fig:rat-wealth-corr} plots Pearson correlations between agents’ traits and payoffs.  Rationality–wealth (blue) and rationality–income (green) climb steeply during the first 75 steps, peak at \(r\!\approx\!0.25\), then taper to \(r\!\approx\!0.18\) as the system stabilizes—showing that cognitive precision yields its greatest returns in the transient learning phase.  Risk aversion moves oppositely: wealth (orange) and income (red) correlations plunge to \(r\!\approx\!-0.07\) and stay negative, implying loss‑averse agents miss early coordination opportunities and remain locked into low-payoff roles.  All curves flatten noticeably once the population converges on near‑harmonious games, so the strongest inequality is created before equilibrium, not after. 

\end{document}